# Self-adaptive waveguide boundary for wideband multi-mode four-wave mixing

*Jianhao Zhang,* [†,§] *Carlos Alonso-Ramos,* [†] *Laurent Vivien,* [†] *Sailing He,* [*,§] *and Eric Cassan* [*,†]

[†] Centre for Nanoscience and Nanotechnology (C2N), CNRS, University Paris-Sud, University Paris-Saclay, 91405 Orsay cedex, France

[§] Centre for Optical and Electromagnetic Research, Zijingang Campus, Zhejiang University, Hangzhou 310058, China

KEYWORDS: Nonlinear optics, phase matching, index manipulation, subwavelength

ABSTRACT: We propose a new approach to provide wideband multi-mode four-wave mixing, independent of the intrinsic waveguide dispersion. We adopt concepts from quantum mechanics and sub-wavelength engineering to design an effective photon well, with a graded potential along the waveguide cross section, that provides flexible control over the mode confinement. The self-adaptive nature of the waveguide boundary allows different spatial modes with equi-spaced frequencies and shared propagation wavevector, automatically fulfilling both, energy conservation and wavevector phase matching conditions. Capitalizing on this concept, we show phase-matching among modes separated by 400 nm (bridging from telecom wavelengths to almost 2**μ**m), with less than 5% deviation in a remarkably large bandwidth exceeding 300 nm. Furthermore, we also show the flexibility of the proposed approach that can be seamlessly adapted to different technology platforms with the same or different waveguide thicknesses. This strategy opens a new design space for versatile nonlinear applications in which the manipulation of energy spacing and phase matching is pivotal, e.g. all-optical signal processing with four-wave mixing, mid-infrared light generation, Brillouin scattering with selectable phonon energy, etc.

**INTRODUCTION**

Nonlinear processes like four-wave mixing (FWM) have garnered a great interest due to their unique capabilities for on-chip light generation, with an immense potential for the implementation of wideband sources for silicon photonics [1-4]. Harnessing 3rd nonlinear Kerr effect in silicon already allowed the demonstration of promising frequency combs [5-8], optical parametric amplification [9-12] and mid-infrared light sources in silicon [13-15]. A great effort has been devoted to compensate both the intrinsic material dispersion and the nonlinearity-induced dispersion, that hamper phase matching which is key to maximize the efficiency of FWM processes [2]. Nevertheless, achieving broadband phase-matching in simple-to-fabricate and fabrication-tolerant silicon waveguides remains an open question. Optimization of transversal dimensions of conventional strip waveguides, provided phase-matching in a relatively narrow wavelength ranges, thus compromising the bandwidth of the nonlinear wavelength conversion processes [9-15]. Broadband phase-matching has been shown based on optimization of high order diffraction terms, e.g. 4th order waveguide dispersion [11], or by implementing rib geometries [12]. Yet, the proposed solutions require complex fabrication processes, with deposition of different materials, or tight control of rib and slab thicknesses. On the other hand, photonic crystal (PhC) waveguides with flexible dispersion properties, high confinement and large group velocities were also considered as a FWM photonic platform [16-17]. However, the bandgap guiding in PhC waveguides results in a limited bandwidth and a high sensitivity to fabrication imperfections that substantially increase propagation loss and distort dispersion properties. An alternative approach to dispersion-engineered waveguides is the use of photonic cavities to effectively enhance the light-matter nonlinear interactions by resonant enhancement and phase-matching among different cavity modes modes. This method was demonstrated in various configurations, including single rings [5-8, 18], coupled rings [19-21], and coupled nanobeam cavities [22, 23]. Concurrently, PhC cavities with parabolic mirror designs were also proposed for nonlinear frequency comb generation [24-27]. Though these demonstrations showed promising performance, their broadband operation is hampered by trade-offs in phase matching and spatial overlaps.

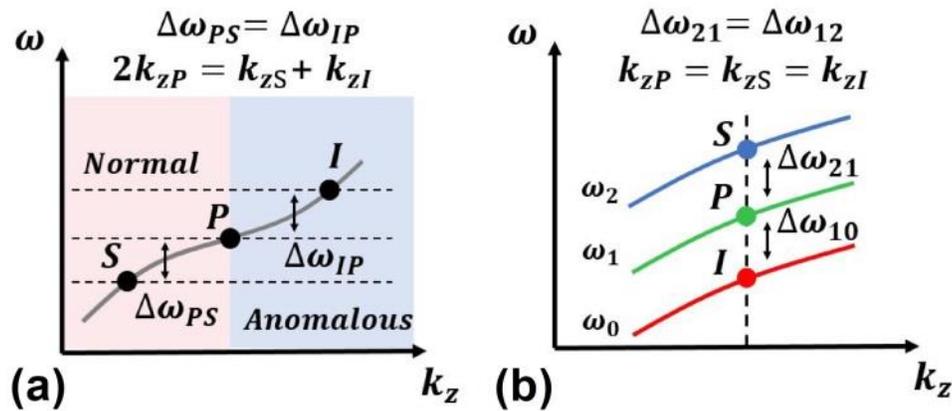

**Fig. 1.** Schematic of degenerate four wave mixing operating in a single-mode waveguide with anomalous dispersion (a) and operating within an inter-modal scheme regardless the absolute

dispersion provided that all dispersion curves are obtained by translating the same curve with a constant frequency step (e.g. $\Delta\omega_{21}=\Delta\omega_{10}$ here).

Here, we propose a new strategy to satisfy both phase matching and energy conservation conditions in an ultra-broad wavelength range. Rather than tuning the waveguide dimensions to yield anomalous dispersion, we shape the index profile of the waveguide to support different spatial modes with the same propagation constant and equal frequency spacing, ensuring phase matching and energy conservation, respectively. In the conventional single-mode FWM approach, depicted in Fig. 1(a), energy is transferred from the pump into signal and idler propagating in the fundamental waveguide mode with different propagation constants. Thus, precise control of dispersion is required to fulfill phase matching condition ($2k_{zP}=k_{zS}+k_{zI}$). In the proposed multi-mode FWM approach, see Fig.1(b), energy is transferred from the pump into signal and idler propagating in different waveguide modes with the same propagation constant ($k_{zP}=k_{zS}=k_{zI}$). Therefore, the phase matching condition is automatically satisfied. Concurrently, energy conservation requires equal frequency spacing. Then, the bandwidth of the proposed scheme does not depend on the exact dispersion of the waveguide, but on the relative slopes of the dispersion curves of the modes, determining the wavelength range where energy conservation is fulfilled. We adopt concepts from quantum mechanics and sub-wavelength engineering to design an effective photon well with a graded potential along the waveguide cross section, yielding a different spatial confinement for each waveguide mode. This self-adaptive boundary simultaneously provides equal frequency spacing, parallel dispersion curves and large spatial overlap among different waveguide modes, overcoming the major bandwidth and efficiency limitations of conventional approaches.

Compared to previous demonstrations using inter-modal FWM [28-31], our work presents a more systematic and controllable strategy with clear analytical explanation. Unlike previous works that relied on parabolic photonic wells along the propagation direction in photonic cavities [27], we exploit a self-adaptive boundary effect in waveguides with a photonic well in the transverse direction to light propagation. This method facilitates broadband wavelength operation compared to the cavity-based four-wave mixing. We also demonstrate that a parabolic guide profile is not essential to obtain a quasi-universal phase matching condition in multimode waveguides with propagative modes. Last, we also show the remarkable flexibility of the proposed approach, allowing large conversion span, with a material-independence design method compatible with sub-wavelength and/or suspended membrane waveguides.

**FREQUENCY SPACINGS OF A STEP-INDEX TWO-DIMENSION (SLAB) WAVEGUIDE**

In order to present the step-by-step concepts, we start here with the simplified case of a slab step-index waveguide, while realistic 3D waveguides corresponding to photonic integration standards are discussed in sections 3 and 4. Let us consider the two-dimension (infinite depth along the x axis) step-index slab waveguide presented in the inset of Fig. 2(a). The eigen equation for modes propagating along the z axis with an electric field polarized along y axis reads as [32]:

$$ha = \frac{m\pi}{2} + arctan(\frac{\gamma n_w^2}{h n_c^2}) \qquad (1)$$

where $n_w$ and $n_c$ are the material index of waveguide core and cladding, while $h = \sqrt{(k_0^2 n_w^2 - k_z^2)}$ and $\gamma = \sqrt{(k_z^2 - k_0^2 n_c^2)}$ are the wavevector along the y axis, inside and outside the waveguide core, respectively. Silicon and silicon dioxide are chosen here as the core and cladding materials, considering material dispersion. The waveguide width $2a$ was set at 700nm which is enough to support 4 modes with effective index values higher than 2. From Equ. (1) we obtain the dispersion curves for the first four order modes (with corresponding mode order $m=0$, 1, 2, 3), shown Fig. 2(a). The frequency spacings change in a nonlinear manner with the mode order $m$, resulting in an uneven spacing that precludes the satisfaction of energy conservation required for the proposed multi-mode FWM approach. Figure 2(b) depicts the evolution of the frequency spacing $\Delta\omega_m = \omega_{m+1}-\omega_m$ and effective index as a function of the mode order. With the increasing mode order m (thus decreasing $n_{eff}$), the frequency spacing $\Delta\omega_m$ monotonously and rapidly increases.

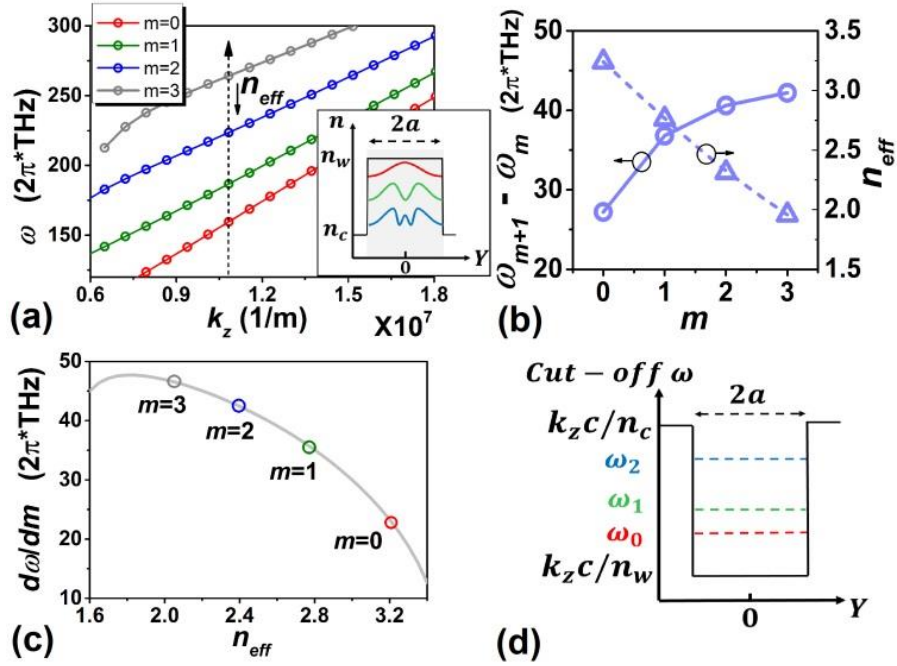

**Fig. 2.** (a) Dispersion curves of first 4 modes of a two-dimension silicon waveguide with silica cladding, propagating along $z$ axis. The width is $2a = 700$ nm. The index profile of the waveguide and mode distribution are plotted in the inset. (b) Frequency spacings $\omega_{m+1}-\omega_m$ and the $n_{eff}$ as a function of the mode order $m$, collected at $k_z = 1.08 \times 10^7$ in (d). (c) frequency

spacings as a function of effective index $n_{eff}$, obtained from analytical calculation. (d) Schematic of a photonic well described by the cut-off frequency for photon propagating along $z$ axis with wavevector $k_z$.

To better understand the evolution of frequency spacing $\Delta\omega_m$, we consider the derivative of the mode frequency with respect to the mode order. The derivative, $d\omega/dm$ in a step-index slab waveguide can be expressed as a function of the core ($n_w$) and cladding ($n_c$) indices, the mode effective index ($n_{eff}$) and the dimension $a$, as (see the Detailed in Supplementary Information S1):

$$\frac{d\omega}{dm} \approx \frac{c\pi\sqrt{n_w^2 - n_{eff}^2}}{2an_w^2} \tag{2}$$

The analytical $d\omega/dm$, calculated from Equ. (2), considering $a$=350nm, $n_w$=3.48, $n_c$=1.445, is plotted in Fig. 2(c). The effective index $n_{eff}$ of the first 4 modes are marked by the circles in the figure. The corresponding values of $d\omega/dm$ range from $2\pi * 24$THz to $2\pi * 45$THz. Marked points are well coincident with the discrete calculation results coming from the dispersion curves shown in Fig. 2(b). From Equ (2) it follows that dm/dω is almost proportional to $\sqrt{n_w^2 - n_{eff}^2}$, which gives a good explanation on the monotonous and nonlinear evolution of $\Delta\omega_m$ observed in Fig. 2(b). This confirms that the analytical frequency spacing $\frac{d\omega}{dm}$ is a useful and simple tool to investigate how the frequency spacing evolves with the waveguide dimension and the index profile.

The step-index slab waveguide can be understood from the point of view of quantum wells, just by considering the cut-off frequencies for the core ($k_zc/n_w$) and cladding ($k_zc/n_c$), for a given wavevector $k_z$ [33-35]. Then, the behavior of frequency spacings is similar to the solutions of harmonic oscillator in a finite-depth potential well. The square potential well formed by the step-index waveguide, depicted in Fig. 2(d), results in unevenly spaced frequencies. Conversely, it is well known that parabolic potential wells yield equi-spaced frequencies [27]. Therefore in the following section we study the influence of the index profile in the frequency separation of the modes.

**FREQUENCY SPACINGS OF A GRADED-INDEX SLAB WAVEGUIDE WITH SELF-ADAPTIVE BOUNDARY**

In this section we study the frequency spacing $\Delta\omega_m$ for a 2D graded-index slab waveguide, with a nonuniform index profile described by $n(y) = (A + By)^p$. By changing the value of $p$, the index profile can be made linear ($p = 1$) or nonlinear ($p \neq 1$). The waveguide width is $2a$, and the indices for the waveguide core, waveguide edge and cladding are $n(0) = n_{cent}, n(a) = n_b$, and $n(|y| > a) = n_c$, as displayed in Fig. 3(a).

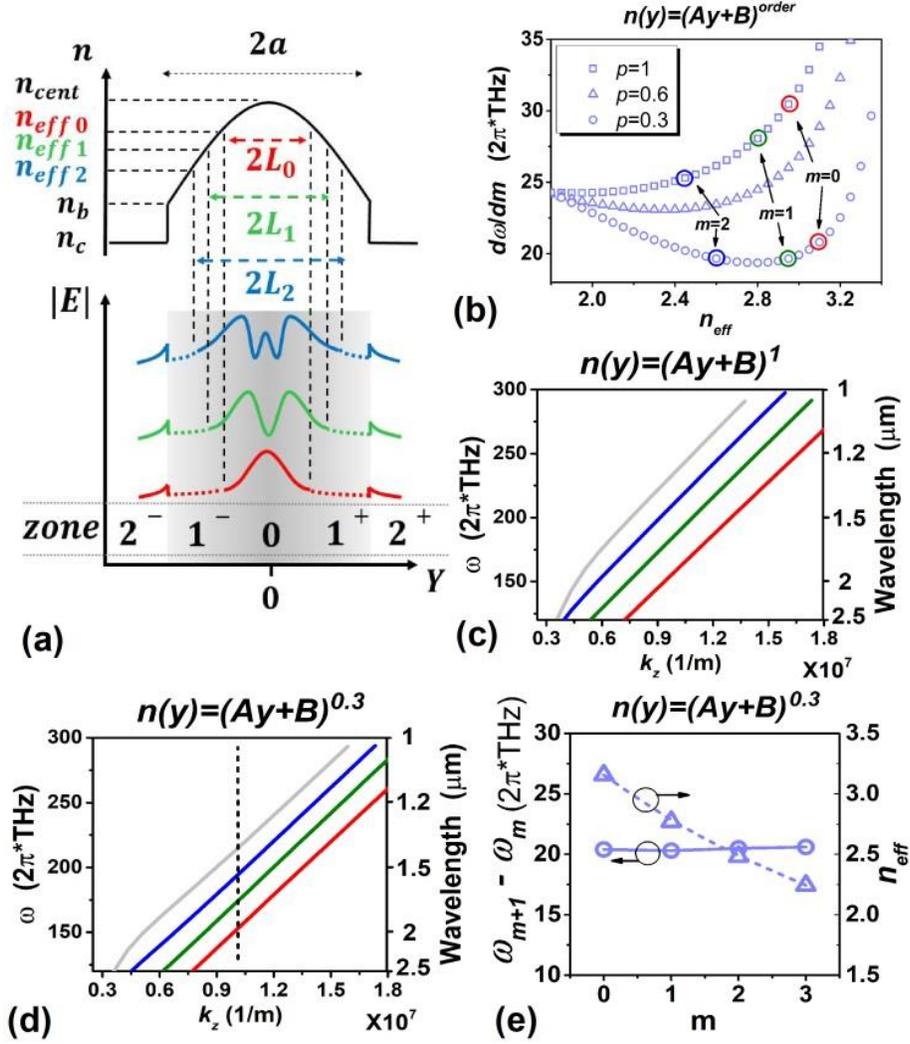

**Fig. 3.** (a) Sketch of a non-uniform index profile and the mode distribution of the first three modes propagating $z$ axis. $n_{cent}$, $n_b, n_c$ are the material index of the waveguide center, waveguide boundary and the surrounding for a waveguide with self-adaptive boundary ($n_{effm}>n_b$). The zero point of $y$ axis is located at the center of the waveguide. (b) Analytical Frequency spacings as a function of effective index $n_{eff}$, with different $order$ numbers (different modes feeling different profiles). The other parameters were adopted as: $2a = 1600$nm. $n_{cent} = 3.48$, $n_b = 1.8, n_c=1.45$. The $n_{eff}$ points, corresponding to the first 3 modes, are labelled by the color circles. (c) Dispersion curves of first 4 modes with $order = 1$, i.e. linear shape, using FDTD

calculation (d) Dispersion curves of first 4 modes with $order = 0.3$, using FDTD calculation. (e) Frequency spacings $\omega_{m+1}$-$\omega_m$ and the $n_{eff}$ as a function of the mode order $m$, collected at $k_z = 1.03 \times 10^7$ in (d).

From the analytical solution of the wave equation for this graded-index waveguide [32] (see Supplementary Information), it follows that the field distribution inside the waveguide core follows a cosinus-like profile while the guide index is larger than the mode effective index ($n(y) > n_{eff}$) and an exponentially decaying profile otherwise ($n(y) < n_{eff}$). An effective mode width, $[-L_m, L_m]$ can be defined as the region where the $n(y) > n_{eff}$, i.e. where mode profile exhibits a cosinus-like shape. Then, as each mode has a different effective index, each mode also has a different effective width. This different confinement effect is illustrated in Fig 3(a), that shows the analytical field profile for the first three waveguide modes. Such a wave confining method, based on the index profile with condition $n_{eff} > n_b$, is then defined as "self-adaptive boundary (SAB)" hereafter. I propose to move the explanations of eq. 3-11 to the supplementary information. First, we consider a linear index profile, i.e. $p = 1$ and $n(y) = Ay + B$. With this profile, the frequency spacings can be described as (see Supplementary Information S2):

$$\frac{d\omega}{dm} = -A\pi c / \left[ n_{eff}^2 \log\left(\sqrt{\frac{n_{cent}^2}{n_{eff,m}^2} - 1} + \frac{n_{cent}}{n_{eff,m}}\right) + n_{cent}\sqrt{n_{cent}^2 - n_{eff,m}^2} \right] \quad (3)$$

Figure 3(b) shows $d\omega/dm$ as a function of $n_{eff}$, calculated from Equ. (3), considering $a$=800nm, $n_{cent}$=3.48, $n_b$=1.8, $n_c$=1.45 and $n_{eff}$ from 1.8 – 3.4. Though the frequency spacings are still changing monotonously, the function exhibits a totally new tendency with an increasing $d\omega/dm$ with $n_{eff}$. This trend is opposite to that of step-index waveguide, shown in Fig. 2(c).

The dispersion curves of the first four modes, obtained from 2D FDTD are presented in Fig. 3 (c), in which the narrowing effect of $d\omega/dm$ is consistent with the prediction shown in Fig. 3(b). Particularly, $d\omega/dm$ is flattened for $n_{eff} \sim 2$, see Fig. 3(b), which indicates that identical frequency spacings can be found at more than one point.

Then, we study $d\omega/dm$ for a nonlinear index profile, with $p \neq 1$ and $n(y) = (A + By)^p$. To do so, we expand $n(y)$ around y=0 to a summation of 8-order polynomial using the Taylor Series Expansion $N(L) = \int_0^L \sum_1^k [C_0 + C_1(n - n_{cent}) + \cdots + C_{10}(n - n_{cent})^k] dy$ and redo the analytical integral and derivative. The evolution of $d\omega/dm$ as a function of $n_{eff}$ for $p = 0.6$ and $p = 0.3$, are presented in Fig. 3 (b). The flattened $d\omega/dm$ region shifts to higher $n_{eff}$ values with decreasing $p$ numbers. For $p = 0.3$ the flattened region appears near $n_{eff} = 2.85$, which is close to that of fundamental mode in a silicon on insulator (SOI) waveguide. Figure 3(d) shows the dispersion curves of the first 4 order modes calculated using FDTD method for $p = 0.3$. Clearly, all the curves are almost parallel and separated by a very close frequency spacing of $\sim 20.5$ THz. The evolution of frequency spacings, $\Delta\omega_m$, and the effective index, $n_{eff}$, for a wavevector of $k_z = 1.03 \times 10^7$ is presented in Fig. 3(e). Opposite to the step-index waveguide studied in Fig. 2(b), the frequency spacing now remains almost constant when changing the mode order $m$, regardless of the sharply descending $n_{eff}$, which is perfectly coincident to the prediction indicated by the analytical calculation. These results illustrate the potential of the proposed approach to achieve energy conservation and phase matching simultaneously, overcoming the major limitations of step-index waveguides.

In the following section we present several realistic implementations of this type of multimode graded waveguides, chosen to illustrate the method. However, we would like to highlight the very general nature of the proposed approach, which can be seamlessly adapted to all types of geometries, photonic platforms (Si, III/V, polymer guides, etc.), and spectral ranges (near infrared, medium infrared).

## GRADED-PROFILE THREE-DIMENSION WAVEGUIDE WITH SELF-ADAPTIVE BOUNDARY

The self-adaptive boundary presented in previous section, which conceptually inherits from the ability of non-uniform potential well, provides equi-spaced frequency modes sharing the same propagation constant, as shown in Fig. 4. One further step is to verify this method in a more practical situation, i.e. with three-dimensional waveguides. As a large index change along the waveguide cross-section is not always feasible from the material point of view, an alternative approach is to implement this graded index profile through nanostructured subwavelength grating (SWG) engineering. By periodically combining high-index and low-index sections with a pitch shorter than half of the wavelength, subwavelength gratings allow the implementation of a material with a synthetic refractive index that can be tuned at will between those of the high and low index materials [36-38]. In Fig. 4(a), the unit cell of two types of periodically-structured subwavelength waveguides are presented for illustration. Here, the gradual index variation is implemented by apodization of the waveguide length or size of the engraved holes. Based on this index-equivalent effective material method and the Marcatili's waveguide approximation, we are again able to semi-analytically investigate the three-dimensional waveguide (Detailed in Supplementary Information S3).

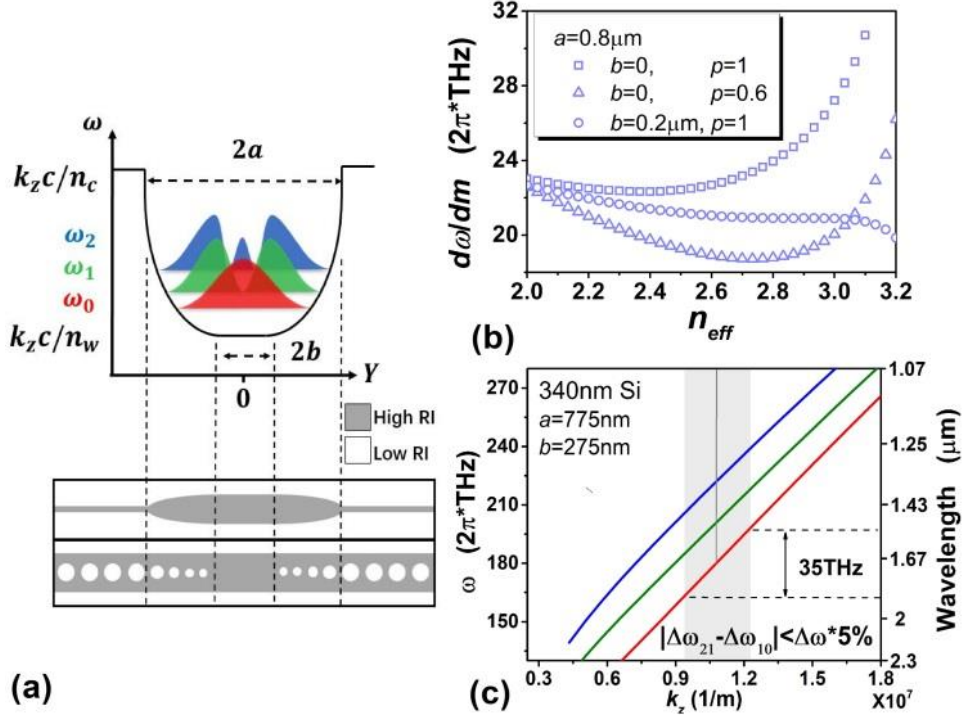

**Fig. 4.** (a) Sketches of a photonic well described by the cut-off frequency for photon propagating along $z$ axis with wavevector $k_z$. The unit cells of two types of graded-structure waveguide is presented in the inset. (b) Analytical Frequency spacings as a function of effective index $n_{eff}$, with different *order* and $b$ values. The other parameters are adopted as: $n_{cent} = 3.48$, $n_b = n_c = 1.8$. (c) Frequency spacings as a function of effective index $n_{eff}$, using 3D FDTD calculation. The perfect matching point and the 5% tolerant range is label by grey line and grey region, respectively.

Calculation is performed on a sub-wavelength waveguide (type in first subfigure of Fig. 4(a)) with a silicon core and air cladding, and nano-arms which are able to support free-standing membrane and provide thermal conduction [39, 40]. The period, thickness and the half waveguide width $a$ are set to 150 nm, 340 nm and 800 nm, respectively. The width of nano-arms is set to 40 nm which gives a boundary index of $n_b = n_c = 1.8$. Particularly, instead of merely adjusting the lineshape of the index profile, we introduce an additional fixed-index section to the center of the waveguide, with a width of [-b,b] to raise the $n_{eff}$. The index between b and a follows the same law $(A + By)^p$ in previous section. Figure 4(b) shows the d$\omega$/d$m$ calculated d$\omega$/d$m$ (using Equ. S20 in Supplementary Information) for waveguides with different parameters. First, we consider a fully graded index profile, with no fixed index region in the center ($b = 0$). For a linear index profile, i.e. $p = 1$, the graded waveguide yields a flattened response near $n_{eff} = 2.4$. With a slower nonlinear index variation, $p = 0.6$, the flattened region can be shifted neff up to 2.75. Further reduction of the p value compresses the flattened region to a narrow region which restricts the available design space. By introducing the fixed index region, we release a new degree of freedom that helps widen the flattened region. By appropriately adjusting the value of $b$ (e. g. to 200 nm), the slope between the local maximum and minimum can be substantially reduced, as shown in Fig. 4(b).

With optimized parameters of $a = 775$nm, $b = 275$nm and $p = 1$, we can locally equalize the frequency spacing and parallelize the dispersion curves. In Fig. 4(c) we show the dispersion curves for the three first modes of the optimized waveguide, calculated using 3D FDTD simulations. With the point $\omega_2 - \omega_1 = \omega_1 - \omega_0$ located at $k_z = 1.07 \times 10^7$, the frequency range with condition $|\Delta\omega_{21} - \Delta\omega_{10}| < \Delta\omega * 5\%$ is as large as 35 THz (~300nm), which also evidences a good tolerance to the possible structure fabrication imperfections. Using this configuration, phase matching is achieved between signal frequency of 220THz (1.36μm) and idle frequency of 172THz (1.72μm). We have to mention that this result came from the operation of first 3 modes as a demonstration. However, this design strategy can be scaled up to higher order modes, e.g. 5 modes. As mentioned earlier, these capabilities arise from the self-adaptive boundary (SAB) condition $n_{eff,m} > n_b, \forall m$ which is more than a simple graded-profile condition. In the Supplementary Material S4, we analytically show that a multimode graded-index waveguide with a classical condition $n_{eff} < n_b$ cannot equalize frequency spacings between its modes since it adopts the same effective boundary for different modes as in a step-index waveguide.

In addition to the demonstrated capability and flexibility of the proposed approach for degenerate four-wave mixing, the nonlinearities-induced phase mismatch can be considered here as well. The phase matching considering nonlinearities is described [2, 11] as $\Delta k = 2\gamma P_P - (2k_{zP} - k_{zS} - k_{zI})$, which is governed by the nonlinear part $2\gamma P_P$ and linear part $\Delta k_L = 2k_{zP} - k_{zS} - k_{zI}$. Since $2\gamma P_P$ is normally positive in silicon, the linear dispersion $\Delta k_L$ needs to be a bit larger than zero to fulfill the global phase matching condition, which is classically addressed by tuning the dispersion to its anomalous regime in a classical waveguide. In contrast, in our case, the condition is translated to the fact that: the frequency spacing $\Delta\omega_{SP}$ should be slightly different from $\Delta\omega_{PI}$, which can be easily achieved just by slightly moving the operating point of pump wave, as illustrated in Fig. 5(a).

If the condition that $\Delta\omega_{SP} = \Delta\omega_{21} - \frac{\delta\omega}{2} = \Delta\omega_{10} + \frac{\delta\omega}{2} = \Delta\omega_{SI}$ can be satisfied at the shifted position characterized by $2k_{zP} > (k_{zS} + k_{zI})$, then the nonlinearity-induced phase mismatch can be compensated, as in Fig. 5(a). This condition requires $\frac{d(\Delta\omega_{21} - \Delta\omega_{10})}{dk_z} > 0$, which can be easily satisfied by adjusting the structure-profile. For example, in the in Fig. 5 (b) we plot the $\delta\omega$ for the optimized waveguide shown in Fig. 4. rightward the working point, we see that $\Delta\omega_{21} > \Delta\omega_{10}$ with increasing $k_z$ (grey line). The negative-to-positive trend well validates the possibility of compensating the nonlinear mismatch. For silicon waveguide working at telecom wavelengths, the effective nonlinear Kerr nonlinearity γ can be described [41, 42] as $\gamma = \frac{\omega n_2 a \Gamma_V}{cV} \cdot \left(\frac{n_g}{n_{Si}}\right)^2$, in which $n_2$, $a$ and $\frac{a\Gamma_V}{V}$ are the nonlinear refractive index, period of SWG structure and effective area of nonlinearity, respectively. By doing the 3D power integral in a single unit cell of the SWG [41, 42], we can obtain an Energy confinement factor (energy confined in silicon) $\Gamma_V$ of 0.87, 0.816 and 0.68 for the idle, pump and signal wave, respectively. With the parameters provided above and $n_2 = 2.8 \times 10^{-18} (m^2/W)$ [43], an effective nonlinear Kerr nonlinearity γ of 65 $(m \cdot W)^{-1}$ and therefore a nonlinear phase item $2\gamma P_P$ of $130\ m^{-1}$ can be deduced. Inter-modal electric-field overlaps are verified as well by calculating the 3D integral $\frac{\iiint E_{yS} \cdot E_{yP} \cdot E_{yP}^* \cdot E_{yI}^* dr^3}{\iiint |E_{yP}|^2 dr^3 \sqrt{\iiint |E_{yS}|^2 dr^3} \cdot \sqrt{\iiint |E_{yI}|^2 dr^3}}$ in a single unit cell [30], which gives a value of 0.12., $E_{yS}$, $E_{yP}$ and $E_{yI}$ corresponding to the field of signal wave, pump wave and idler waves.

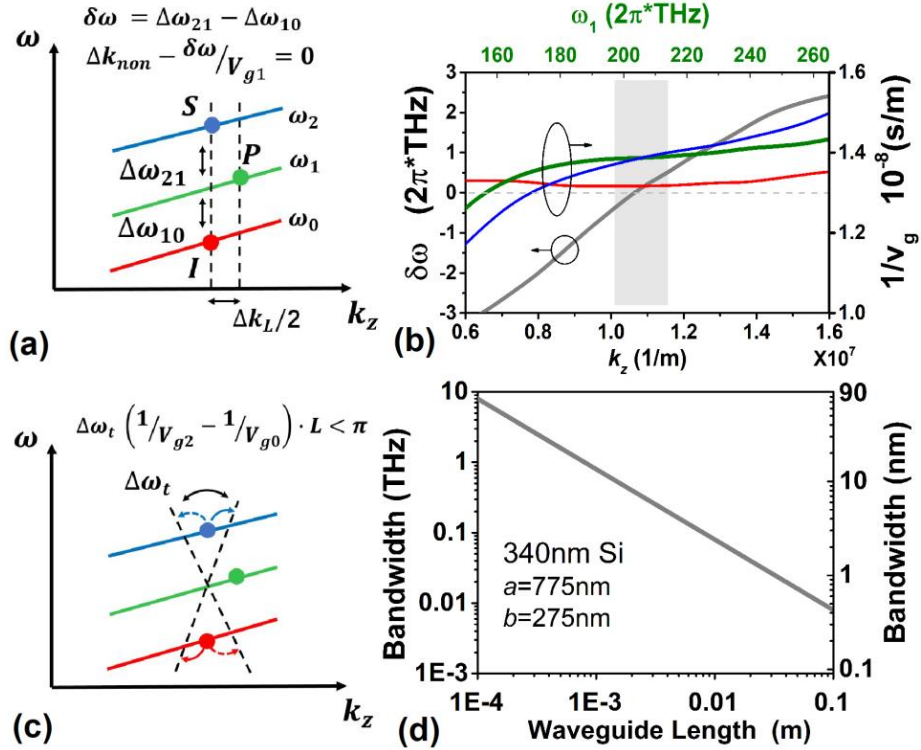

**Fig. 5.** (a) Schematic of the proposed SWG waveguide for the simultaneous energy conservation and wavevector phase matching under the effect of a nonlinear effect. (b) The difference between frequency spacings $\delta\omega = \Delta\omega_{21} - \Delta\omega_{10}$ as a function of wavevector is corresponding to the left axis. On the right side arethe reciprocal of group velocity of the first 3 modes. The working range is labelled by the gray region. (c) Schematic of using the proposed SWG waveguide for tunable four-wave mixing (d) The tunable band width as a function of the waveguide length.

An important consideration is to discuss the tunability operation of the designed multimode waveguide, for which $1/V_g$, i. e. $\frac{dk_z}{d\omega}$, of the first 3 modes are shown in Fig. 5(c) and Fig. 5(b), Vg being the mode group velocity. As it is well known, the 3dB bandwidth of the FWM process can be described as $|\Delta k| L_{wg} = \Delta\omega_t (\frac{1}{V_{g2}} - \frac{1}{V_{g0}}) L_{wg} < \pi$, in which $L$ is the waveguide length. With a value of $0.07 \times 10^{-8}$ (s/m) for $(\frac{1}{V_{g2}} - \frac{1}{V_{g0}})$ obtained from Fig. 5(b), we can predict an 3dB tunable bandwidth of around 1THz (~10nm) for a 1mm long waveguide. $V_{g2}$ and $V_{g0}$ are corresponding to the 2nd and fundamental modes, respectively By generalizing this approach, the 3dB tunable bandwidth as a function of waveguide length can be plotted in Fig. 5(d). It can therefore be seen that, even in a non-optimized waveguide, the spectral operating band of the FWM process is as wide as few tens of nanometers for sub-millimeter long waveguide.

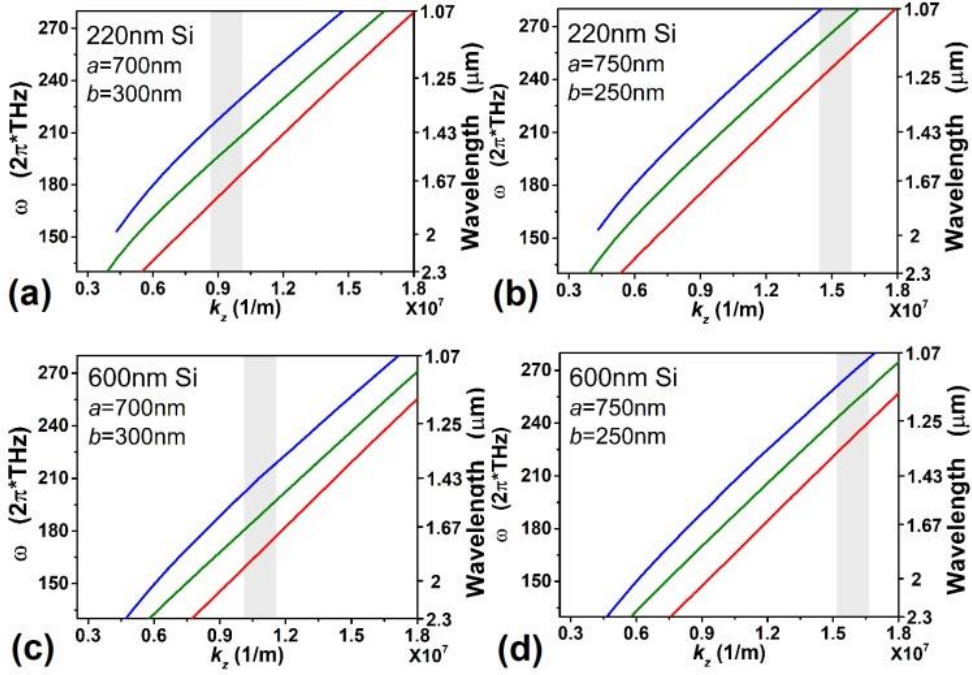

**Fig. 6.** (a), (b), (c) and (d) are the dispersion curves of first 3 modes in SWG waveguide with different lengths $a$, $b$ and linear strategy ($order = 1$). The working points at which energy conservation and phase matching is satisfied, is labelled by grey region.

With all these results, we have successfully demonstrated the feasibility of using a multimode SWG waveguide with self-adaptive boundary for degenerate FWM, based on the interaction among the first three modes. However, nonlinear processes are not strictly limited to this scenario. Using higher order modes (e. g. fundamental mode, 2nd mode and 4th mode), FWM process could be further scaled to even larger conversion spans by using the same method, for example to directly convert near-infrared waves to nearly mid-infrared waves. Most importantly, since the proposed approach no longer requires specific normal/anomalous dispersion, we can easily adopt this approach to any other platforms, no matter what kinds of dispersion the waveguide modes exhibit.

As an illustrative example of the highly adaptable nature of the method to different platforms with different thicknesses, targeting different frequency ranges, etc, we investigated its application to different situations. Using, 3D FDTD method and considering material dispersion, we calculated the dispersion curves for Si sub-wavelength waveguides with Si thicknesses ranging between 220nm and 600nm. For the 220nm-thick Si waveguide with parameters $a$=700nm, $b$=300nm, the optimized position for $\Delta\omega_{21}=\Delta\omega_{10}$ is located at wavevector around $k_z = 0.95 \times 10^7$ m-1, leading to a conversion span from 220 THz to 180 THz, as shown in Fig. 6(a). By raising $a$ and reducing $b$ to 750nm and 250 nm, we have been able to shift the optimized position continuously to $k_z = 1.53 \times 10^7$ m-1, with starting and stop frequencies of 280 THz to 245 THz, respectively. In order to push the operation point to a lower frequency, with the purpose of generating light wavelengths up to $2\mu m$, larger waveguide cross-sections can be considered. Very interestingly, by simply enlarging the thickness to 600nm, the working frequency for the fundamental mode can be shifted to around 155THz, with conversion span over 40THz (from 1.53µm to 1.93µm, i.e. 400 nm), as shown in Fig. 6(c), with almost no displacement on $k_z$. Similar result can also be observed in another configuration ($a$=750nm, $b$=250nm), with a slightly shifted working point $k_z$. These gathered results unambiguously show the simplicity and flexibility of our approach for degenerate four-wave mixing, i. e., 1) for each waveguide thickness the waveguide is capable to offer adjustable working conditions in a wide frequency range, within a varying index-profile; 2) for a certain optimal lineshape, the strategy for shifting the working wavelength, is to simply adjust the thickness.

**CONCLUSION**

With thorough analytical explanation and numerical confirmation, we show that waveguides with material-engineered graded index profile (unambiguous condition that $n_{eff} > n_b$) can be designed to support the modes that adapt themselves to different effective boundaries ("self-adaptive boundary"), according to their effective index values. This self-adaptive behavior, that relies on similar concepts as those found for potential wells, provides new degrees of freedom to achieve simultaneous satisfaction of both the energy conservation and phase matching, regardless the intrinsic dispersion of the considered optical waveguide modes. This strategy, that can be adapted to different wavelength ranges and material platforms, opens a new design space for degenerate FWM. Capitalizing on this concept, we show that phase-matching condition can be satisfied over 400nm (bridging from telecom wavelength to almost 2μm) by employing three waveguide modes. This range could be further expanded by using higher order modes enabled by flexible index profile optimization. We foresee that the self-adaptive boundary concept will expedite the development of a new generation of nonlinear circuits with an immense potential for light generation in mid-infrared wavelength, but also for nonlinear process beyond FWM and applications in which dispersion manipulation is of major relevance.

**SUPPLEMENTARY MATERIAL**

S1 Calculation for a step-index two-dimensional waveguide; S2 Calculation for a graded-index two-dimension waveguide with self-adaptive boundary ($n_{eff} > n_b$); S3 Semi-analytical calculation for a graded-profile three-dimensional waveguide with self-adaptive boundary ($n_{eff} > n_b$); S4 Discussion on a graded-index waveguide with classical condition $n_{eff} < n_b$.

**ACKNOWLEDGMENT**

The French ANR agency is acknowledged for the its support through the SITQOM project. The European Research Council (ERC) is acknowledged for its support through the POPSTAR project. National Natural Science Foundation of China (NSFC) (61774131) is also acknowledged.

# Supplementary materials

**S1. CALCULATING THE $d\omega/dm$ OF A STEP-INDEX TWO-DIMENSION WAVEGUIDE**

As introduced in the text, for a two-dimension (infinite depth along $x$ axis) step-index waveguide, the eigen equation for modes propagating along $z$ axis with electric field polarizing along $y$ axis can be described as $ha = \frac{m\pi}{2} + arctan\left(\frac{\gamma n_w^2}{h n_c^2}\right)$, in which $n_w$ and $n_c$ is the material index of waveguide core and cladding while $h = \sqrt{(k_0^2 n_w^2 - k_z^2)}$ and $\gamma = \sqrt{(k_z^2 - k_0^2 n_c^2)}$ are the wavevector along the $y$ axis, inside and outside the waveguide, respectively. Assuming that the $k_z$ is a constant and $\omega$ is a function of $n_{eff}$ and $m$, the frequency spacing between modes can therefore be expressed as:

$$\frac{d\omega}{dm} = \frac{d}{dm}\left(\frac{k_z c}{n_{eff}}\right) = \frac{k_z c}{-n_{eff}^2} \cdot \frac{dn_{eff}}{dm} = \frac{k_z c}{-n_{eff}^2} \bigg/ \frac{dm}{dn_{eff}} \tag{S1}$$

From Equ. (S1) we can infer that if $dm/dn_{eff}$ is proportional to $-1/n_{eff}^2$, then the frequency spacing is fixed. Though the mode order number $m$ is a discrete integer indicating the phase solution, mathematically it still can be represented as a function of $\omega$ and $n_{eff}$, from Equ. (1) that:

$$m = \frac{2}{\pi}\left[\frac{k_z \cdot a \sqrt{n_w^2 - n_{eff}^2}}{n_{eff}} - arctan(\frac{\gamma n_w^2}{h n_c^2})\right] = f_{m1} + f_{m2} \tag{S2}$$

In which $f_{m1} = \frac{2k_z}{\pi} a \sqrt{\frac{n_b^2}{n_{eff}^2} - 1}$ describing the phase from standing waves, while $f_{m2} = -\frac{2}{\pi} arctan(\frac{n_b^2 \sqrt{(n_{eff}^2 - n_c^2)}}{n_c^2 \sqrt{(n_b^2 - n_{eff}^2)}})$ accounts for the abrupt index change on the boundary. The related first derivatives can be written as:

$$\frac{df_{m1}}{dn_{eff}} = \frac{2k_z}{\pi} \frac{d}{dn_{eff}} \left( a\sqrt{\frac{n_w^2}{n_{eff}^2} - 1} \right) = \frac{-2ak_z n_w^2}{\pi n_{eff}^2 \sqrt{n_w^2 - n_{eff}^2}} \tag{S3}$$

$$\frac{df_{m2}}{dn_{eff}} = -\frac{2}{\pi} \cdot \frac{d}{dn_{eff}} arctan(\frac{\gamma n_w^2}{h n_c^2}) = -\frac{2}{\pi} \cdot \frac{1}{1 + \frac{n_b^4(n_{eff}^2 - n_c^2)}{n_c^4(n_b^2 - n_{eff}^2)}} \cdot \frac{d}{dn_{eff}} \left( \frac{n_b^2 \sqrt{(n_{eff}^2 - n_c^2)}}{n_c^2 \sqrt{(n_b^2 - n_{eff}^2)}} \right)$$

$$= -\frac{2}{\pi} \cdot \frac{n_{eff}}{-1 + n_{eff}^2(1/n_c^2 + 1/n_w^2)} \cdot \frac{1}{\sqrt{(n_{eff}^2 - n_c^2)}} \cdot \frac{1}{\sqrt{(n_w^2 - n_{eff}^2)}} \tag{S4}$$

Confirmed by the numerical methods, for $a$=350nm, $n_w$=3.48, $n_c$=1.445 and $n_{eff}$ from 1.6 to 3, we have $\frac{df_{m2}}{dn_{eff}} \in (-1.7, -0.19)$ while $\frac{df_{m1}}{dn_{eff}} \in (-5, -3)$, from which we may assume that the $\frac{dm}{dn_{eff}}$ is almost governed by $\frac{df_{m1}}{dn_{eff}}$ for simplifying the analysis. Therefore, $\frac{dm}{dn_{eff}}$ can be approximately expressed as:

$$\frac{d\omega}{dm} = \frac{k_z c}{-n_{eff}^2} / (\frac{df_{m1}}{dn_{eff}} + \frac{df_{m2}}{dn_{eff}}) \approx \frac{k_z c}{-n_{eff}^2} / (\frac{df_{m1}}{dn_{eff}}) = \frac{c\pi \sqrt{n_w^2 - n_{eff}^2}}{2an_w^2} \tag{S5}$$

## S2. CALCULATING THE $d\omega/dm$ OF A GRADED-INDEX TWO-DIMENSION WAVEGUIDE WITH SELF-ADAPTIVE BOUNDARY ($n_{eff} > n_b$)

We consider a 2D graded-index slab waveguide (then with graded cut-off potential) with a waveguide width of $2a$. The indices for the waveguide core, waveguide edge and cladding are $n(0) = n_{cent}$, $n(a) = n_b$ and $n_c$, respectively, as displayed in Fig. S1. we propose to systematically rely on a nonuniform index profile $n(y)$. When we introduce a condition that the effective indices of the modes are larger than the physical boundary index (i.e. $n_{eff} > n(a) = n_b$ for all the guided modes) into the index profile, the waveguide can be considered as splitting into five zones (zones 0, ∓1, ∓2 in the Fig. S1). The central part (part 0) is within the $[-L_m, L_m]$ range in which $n(y)$ is larger than $n_{eff}$ and can be expressed by a $cos$ function as usually, instead of the physical boundaries (i. e. $y = \mp a$). The other four zones (zones ∓1, ∓2) are described by a decaying form because $n(y)$ is there smaller than $n_{eff}$.

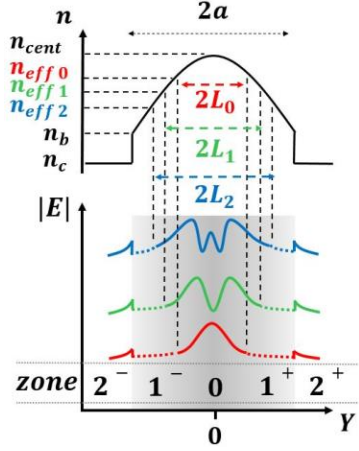

**Fig. S1.** (a) Sketch of a non-uniform index profile and the mode distribution of the first three modes propagating $z$ axis. $n_{cent}$, $n_b, n_c$ are the material index of the waveguide center, waveguide boundary and the surrounding for a waveguide with self-adaptive boundary ($n_{effm} > n_b$). The zero point of $y$ axis is located at the center of the waveguide.

To satisfy the Maxwell's equations and the corresponding boundary condition for an electric field polarizing along $y$, the field can be expanded as follows we rewrite the wave equation inside the waveguide [1]:

$$H_x = A_2^+ exp\left[k_0\sqrt{n_{eff}^2 - n_c^2} \cdot (y - a)\right], \quad a \leq y \tag{S6}$$

$$H_x = A_1^+ cosh[D(y)] + B_1^+ sinh[D(y)], \quad L \leq y < a \tag{S7}$$

$$H_x = A_0 cos[N(y) - \varphi], \quad -L < y < L \tag{S8}$$

$$H_x = A_1^- \cosh[D(y)] + B_1^- \sinh[D(y)], \quad -a < y \leq L \tag{S9}$$

$$H_x = A_2^- \exp\left[k_0\sqrt{n_{eff}^2 - n_c^2} \cdot (y+a)\right], \quad -a \geq y \tag{S10}$$

in which the $A_{1,2}^+$ and $B_{1,2}^+$ are the amplitudes of the decaying components for the positive direction, while $A_{1,2}^-$ and $B_{1,2}^-$ stand for the negative direction. $\varphi$ is the biased phase that is related to the mode order ($\varphi = 0$ or $\varphi = \pi/2$, with respect to modes of symmetric and anti-symmetric parities). The pattern of zone 0 and zones $\mp 2$ jointly determine the hyperbolic form of part 1. Due to the varying index profile, currently the total phases of the wave propagation along the $y$ axis inside the waveguide can be transformed to the integrals of position-related wavevector $\sqrt{(k_0^2 n^2(y) - k_z^2)}$ to $y$ coordinate ($N_m$ and $D_m$ stand with respect to zone 0 and zones $\mp 1$, respectively):

$$N_m(y) = \int_0^{L_m} \sqrt{(k_0^2 n^2(y) - k_z^2)}\, dy \tag{S11}$$

$$D_m(y) = \int_{L_m}^y \sqrt{(k_z^2 - k_0^2 n^2(y))}\, dy \tag{S12}$$

As it is shown in Fig. S1, modes with different mode orders m are confined in different spatial spans, which sizes increase with m. Each spatial span $L_m$ (for m=1,2,3, …) which can be called "effective length" for mode 'm' is defined through $n(L_m) = n_{effm}$. The condition $\boldsymbol{n_{eff} > n_b}$ is defined as "self-adaptive boundary (SAB)" in the manuscript. By solving the equation (S6)-(S12), the eigen equation can be written as:

$$\tan[N(L_m) \pm \varphi] = \frac{D_m'(L_m)}{N_m'(L_m)} = \lim_{y \to L_m} \frac{D_m'(y)}{N_m'(y)} = 1 \tag{S13}$$

The Eigen equation for modes propagating along z axis with an electric field polarized along the y axis can be rewritten as:

$$N(L_m) = \int_0^{L_m} \sqrt{(k_0^2 n^2(y) - k_z^2)}\, dy = \frac{m\pi}{2} + \frac{\pi}{4} \tag{S14}$$

Previously, in the step-index waveguide, to respond to the approximately linear phase increas on the right part of Equ. (1), the changes of $\omega$ and $n_{eff}$ were correlated thorugh the relationship $a\left(\frac{\omega_{m+1}}{c}\right)\sqrt{n_w^2 - n_{eff,m+1}^2} - a\left(\frac{\omega_m}{c}\right)\sqrt{n_w^2 - n_{eff,m}^2} = \frac{\pi}{2}$. When the proposed self-adaptive boundary $n_{eff,m} > n_b\ \forall m$ is introduced, the spatial integral range is automatically selected which exactly gives the room to trim the frequency spacing through the following condition :

$\frac{\omega_{m+1}}{c}\int_0^{L_{m+1}}\sqrt{n^2(y) - n_{eff,m+1}^2} - \frac{\omega_m}{c}\int_0^{L_m}\sqrt{n^2(y) - n_{eff,m}^2} = \frac{\pi}{2}$ .To consider the improvement from this new condition, a similar analysis for frequency spacings is carried out as previously with a general index profile described by $n(y) = (A + By)^p$. First considering a linear index profile, i.e $n(y) = Ay + B$ the phase $N(L_m)$ can be recalculated as:

$$N(L_m) = \frac{m\pi}{2} + \frac{\pi}{4} = \frac{k_z n(y)}{2An_{eff,m}}\sqrt{n^2(y) - n_{eff,m}^2} - \frac{k_z n_{eff,m}}{2A}\log\left[\frac{k_z}{n_{eff,m}}\sqrt{n^2(y) - n_{eff,m}^2} + \frac{k_z}{n_{eff,m}}n(y)\right]\Big|_0^L \tag{S15}$$

Because of the self-adaptive behavior that the endpoint $n(L_m)$ is equal to $n_{effm}$, then the (S15) integral can be greatly simplified in a common form for different modes, with $n(0) = B = n_{cent}$, to:

$$N(L_m) = \frac{-k_z n_{eff,m}}{2A}\log[k_z] - \left(\frac{n_{cent}}{2A}\cdot\frac{k_z}{n_{eff,m}}\sqrt{n_{cent}^2 - n_{eff,m}^2} - \frac{k_z n_{eff,m}}{2A}\log\left[\frac{k_z}{n_{eff,m}}\sqrt{n_{cent}^2 - n_{eff,m}^2} + \frac{k_z}{n_{eff}}n_{cent}\right]\right)$$

$$= \frac{k_z n_{eff,m}}{2A}\log\left(\sqrt{\frac{n_{cent}^2}{n_{eff,m}^2} - 1} + \frac{n_{cent}}{n_{eff,m}}\right) - \frac{k_z n_{cent}}{2A}\sqrt{\frac{n_{cent}^2}{n_{eff,m}^2} - 1} \tag{S16}$$

By doing the 1st derivative of $N(L)$, we have

$$\frac{dN(L_m)}{dn_{eff}} = \frac{k_z}{2A} \left( \log\left[\sqrt{\frac{n_{cent}^2}{n_{eff,m}^2} - 1} + \frac{n_{cent}}{n_{eff,m}}\right] - \frac{\frac{n_{eff,m}^2}{\sqrt{n_{cent}^2 - n_{eff,m}^2}}}{\sqrt{n_{cent}^2 - n_{eff,m}^2} + n_{cent}} - 1 + \frac{n_{cent}^3}{n_{eff,m}^2 \sqrt{n_{cent}^2 - n_{eff,m}^2}} \right)$$

$$= \frac{k_z}{2A} \log\left(\sqrt{\frac{n_{cent}^2}{n_{eff,m}^2} - 1} + \frac{n_{cent}}{n_{eff,m}}\right) + \frac{k_z n_{cent} \sqrt{n_{cent}^2 - n_{eff,m}^2}}{2A n_{eff,m}^2} \quad (S17)$$

For a fixed $k_z$, we have $m = \frac{2}{\pi}\left(N(L_m) - \frac{\pi}{4}\right)$, therefore

$$\frac{d\omega}{dm} = \frac{k_z c}{-n_{eff,m}^2} \Big/ \frac{dm}{dn_{eff,m}} = \frac{\pi k_z c}{-2n_{eff,m}^2} \Big/ \frac{dN(L)}{dn_{eff,m}}$$

$$= -A\pi c / [n_{eff,m}^2 \log\left(\sqrt{\frac{n_{cent}^2}{n_{eff,m}^2} - 1} + \frac{n_{cent}}{n_{eff,m}}\right) + n_{cent}\sqrt{n_{cent}^2 - n_{eff,m}^2}] \quad (S18)$$

Comparing Equ. (S18) to (S5), we see that the previous variable with a square root can be replaced by the reciprocal of a new-introduced combination of logarithmic and square root item of $n_{eff}$. Currently, the waveguide properties is enabled by a new freedom to engineer the frequency spacings.

## S3. CALCULATING THE $d\omega/dm$ OF A GRADED-PROFILE THREE-DIMENSION WAVEGUIDE WITH SELF-ADAPTIVE BOUNDARY ($n_{eff} > n_b$)

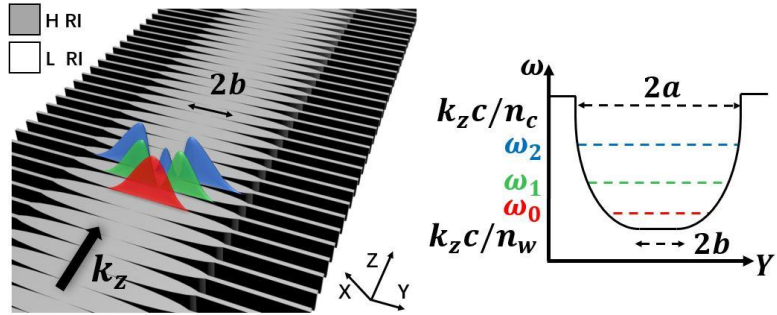

**Fig. S2.** Schematic of a graded-profile subwavelength waveguide. The equivalent cut-off frequency distribution indicated by the index, is presented in the inset.

As an example of the possible design of three-dimensional graded-profile waveguides with a self-adaptive boundary, a subwavelength grating (SWG) waveguide with a period of 150nm is chosen, which the geometry is depicted in Fig. S2. The material considered here are silicon and air, for the high-index and low-index regions, respectively. As the SWG is able to be considered as being made of equivalent materials, the strategy here to achieve the graded-profile (graded-index) waveguide is tapering the component width of silicon, from the center to the edge, while keeping the period fixed. In this case, the width of silicon components is changed from 150nm to 40nm, i. e. adjusting the filling factor from 1 to 0.267. According to the well-known equivalent properties of SWG waveguide, the index of SWG waveguide can be described as:

$$n_{SWG}(y) = \sqrt{\eta n_{cent}^2 + (1-\eta)n_c^2} = \sqrt{\eta(n_{cent}^2 - n_c^2) + n_c^2} \tag{S19}$$

which leads to an index ranging from around 3.48 to 1.8 for the C-band wavelengths. The $\eta(y) = (Ay + B)^{order}$ is the filling factor of silicon of the SWG waveguide at a transverse position $y$. The schematic of potential well that describes e cut-off frequency, as introduced before, is presented in the inset of Fig. S2. In order to approximately investigate the frequency spacings of this three-dimension (3D) SWG waveguide, the Marcatili's method [1] is adopted to find an equivalent simpler 2D waveguide with eight cladding regions, then the phase item of the dispersion modes Equ. (S14) can be rewritten as:

$$N(L_m) = \int_0^{L_m} \sqrt{(k_0^2 n_{SWG}^2(y) - k_x^2 - k_z^2)}\, dy = \frac{m\pi}{2} + \frac{\pi}{4} \tag{S20}$$

$$k_x t = \arctan\left(\frac{\gamma_x}{k_x}\right) \tag{S21}$$

$$\gamma_x = \sqrt{k_0^2 n_{SWG}^2(y) - k_x^2 - k_0^2 n_c^2} \qquad (S22)$$

in which $k_x$ and $\gamma_x$ are the wavevectors along the x direction, inside and outside the waveguide, respectively. The corresponding index component can be depicted by $n_x = \frac{k_x}{k_0}$. Limited by the complicated expression of index of SWG waveguide and the non-analytical solution of the dispersion equation along $y$, finding an accurate solution of (S20) is difficult. Instead, using numerical fitting we simplify the $n_x$ to a function $n_x^2 = C + D n_{SWG}^2(y)$ by numerically solving (S21) and (S22). The comparison between the analytical/numerical solving and the approximation, within the range of $n_{SWG} \in [1.8, 3.48]$, is shown in Fig. S3.

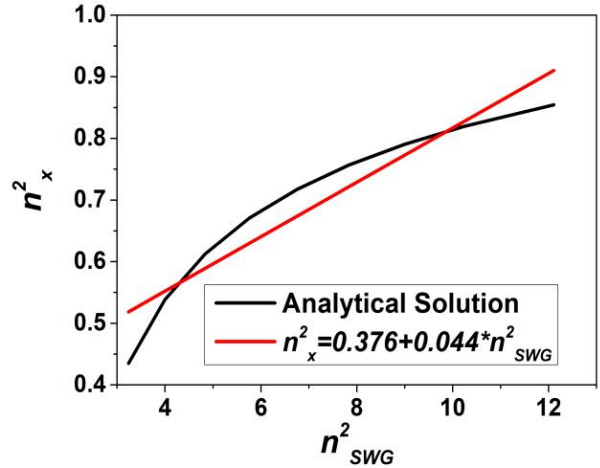

**Fig. S3.** Comparison between the analytical solution and the linear approximation of $n_x^2$, using $n_x^2 = C + D n_{SWG}^2(y)$. The values $C$ and $D$ are directly shown in the figure.

Based on this approximation, this case can be simplified as a two-dimension waveguide with an equivalent "material index" of:

$$n(y) = \sqrt{n_{SWG}^2(y) - n_x^2} = \sqrt{(1-D) n_{SWG}^2(y) - C}$$

$$= \sqrt{(1-D)[\eta(n_{cent}^2 - n_c^2) + n_c^2] - C} \tag{S23}$$

With this equivalent index, (S12) can be simplified to the general form:

$$N(L_m) = \int_0^L \sqrt{(k_0^2 n^2(y) - k_z^2)} \, dy = \frac{m\pi}{2} + \frac{\pi}{4} \tag{S24}$$

Now the phase item shares the same expression as (S14). Thus, we can adopt the method discussed before to optimize the waveguide. Very importantly, the effective index $n_{eff}$ is shifted to a lower level due to the limited thickness, which gives a maximum equivalent index of $n(0) = \sqrt{(1-D)[(n_{cent}^2 - n_c^2) + n_c^2] - C}$.

In order to ease the effect coming from $n_x$ and simultaneously promote the effective index values of all the modes, we introduce additionally a shared section in which the index is fixed as the central section shown in Fig. S2, with a width of $2b$. By setting this, a minimum phase can be ensured and the total phase can be rewritten as:

$$N(L_m) = b \frac{\omega}{c} \sqrt{n^2(0) - n_{eff,m}^2} + \int_b^L \frac{\omega}{c} \sqrt{n^2(y) - n_{eff,m}^2} \, dy \tag{S25}$$

Since $n(L_m) = n_{eff,m}$, if order = 1, that is to say $\eta(y) = (Ay + B)$, then

$$N(L_m) = bk_z \sqrt{\frac{n^2(0)}{n_{eff,m}^2} - 1} + \frac{-2k_z}{3An_{eff,m}(1-D)(n_{cent}^2 - n_c^2)} [n^2(0) - n_{eff,m}^2]^{3/2} \tag{S26}$$

Therefore, the 1st derivative of $N(L)$ is modified as:

$$\frac{dN(L_m)}{dn_{eff}} = -\frac{b\beta n^2(0)}{n_{eff}^2 \sqrt{n^2(0) - n_{eff,m}^2}} + \frac{2\beta}{An_{eff,m}^2(1-D)(n_{cent}^2 - n_c^2)}$$

$$\cdot \left( \frac{[n^2(0) - n_{eff,m}^2]^{3/2}}{3} + n_{eff,m}^2 \sqrt{n^2(0) - n_{eff,m}^2} \right)$$

$$= \frac{2\beta}{An_{eff,m}^2(1-D)(n_{cent}^2-n_c^2)} \left( \frac{\left[n^2(0)-n_{eff,m}^2\right]^{3/2}}{3} + n_{eff,m}^2\sqrt{n^2(0)-n_{eff,m}^2} - \frac{bAn^2(0)(1-D)(n_{cent}^2-n_c^2)}{2\sqrt{n^2(0)-n_{eff,m}^2}} \right)$$

(S27)

We then obtain similarly the frequency spacings as:

$$\frac{d\omega}{dm} = = \frac{-A\pi c(1-D)(n_{cent}^2-n_c^2)}{4} \bigg/ \left( \frac{\left[n^2(0)-n_{eff,m}^2\right]^{3/2}}{3} + n_{eff,m}^2\sqrt{n^2(0)-n_{eff,m}^2} - \frac{bAn^2(0)(1-D)(n_{cent}^2-n_c^2)}{2\sqrt{n^2(0)-n_{eff,m}^2}} \right)$$

(S28)

Currently, the frequency spacing $\frac{d\omega}{dm}$ is no longer simply dominated by the effective index $n_{eff,m}$. Instead, also being affected by the new introduced item $\frac{bAn^2(0)(1-D)(n_{cent}^2-n_c^2)}{2\sqrt{n^2(0)-n_{eff,m}^2}}$, which gives us a new room to control the frequency spacing by providing additionally a local maximum-peak at the high $n_{eff}$ range, as shown in Fig. 4(b) in the main text.

## S4. DISCUSSION ON A GRADED-INDEX TWO-DIMENSION WAVEGUIDE WITH $n_{eff} < n_b$

With the classical condition that ($n_{eff} < n_b$), the eigen dispersion equation can be similarly written by replacing the $\frac{\pi}{4}$ with the boundary-induced phase item $arctan(\frac{n_b^2\sqrt{(k_z^2-k_0^2n_c^2)}}{n_c^2\sqrt{(k_0^2n_b^2-k_z^2)}})$, as:

$$N(a) = \int_0^a \sqrt{(k_0^2 n^2(y) - \beta^2)}\, dy = \frac{m\pi}{2} + arctan(\frac{n_b^2\sqrt{(k_z^2-k_0^2n_c^2)}}{n_c^2\sqrt{(k_0^2n_b^2-k_z^2)}})$$

(S29)

$$m = \frac{2}{\pi}N(a) - \frac{2}{\pi}arctan(\frac{\gamma n_w^2}{hn_c^2}) = \frac{2}{\pi}N(a) + f_{m2}$$

(S30)

If $n(y) = Ay + B$, that is to say, for a linear-index profile waveguide, then

$$N(a) = \left(\frac{(Ay+B)}{2A} \cdot \frac{\omega}{c}\sqrt{(Ay+B)^2 - n_{eff,m}^2} - \frac{k_z n_{eff,m}}{2A}log[\frac{\omega}{c}\sqrt{(Ay+B)^2 - n_{eff,m}^2} + \frac{\omega}{c}(Ay+B)]\right)\Big|_0^a \quad (S31)$$

$$N(a) = \frac{k_z}{2A}\left(n_b\sqrt{\frac{n_b^2}{n_{eff,m}^2} - 1} - n_{cent}\sqrt{\frac{n_{cent}^2}{n_{eff,m}^2} - 1} + n_{eff,m}\log\left[\frac{\sqrt{n_{cent}^2 - n_{eff,m}^2} + n_{cent}}{\sqrt{n_b^2 - n_{eff,m}^2} + n_b}\right]\right)$$
(S32)

Similarly,

$$\frac{d\omega}{dm} = \frac{k_z c}{-n_{eff,m}^2} \Big/ \frac{d}{dn_{eff,m}}[\frac{2}{\pi}N(a) + f_{m2}] \quad (S33)$$

$$= \frac{\pi k_z c}{-2n_{eff,m}^2} \Bigg/ \Bigg\{ \frac{\beta k_z}{2A}\left[\frac{n_{cent}\sqrt{n_{cent}^2 - n_{eff,m}^2}}{n_{eff,m}^2} - \frac{n_b\sqrt{n_b^2 - n_{eff,m}^2}}{n_{eff,m}^2} + \log\left(\frac{\sqrt{n_{cent}^2 - n_{eff,m}^2} + n_{cent}}{\sqrt{n_b^2 - n_{eff,m}^2} + n_b}\right)\right] +$$

$$\frac{-n_{eff,m}}{-1 + n_{eff,m}^2(1/n_c^2 + 1/n_b^2)} \cdot \frac{1}{\sqrt{(n_{eff,m}^2 - n_c^2)}} \cdot \frac{1}{\sqrt{(n_b^2 - n_{eff,m}^2)}} \Bigg\} \quad (S34)$$

Comparing Equ. (S34) to Equ. (S5), the only change we can clearly see is the replacing from $\frac{d}{dn_{eff}}f_{m1}$ to $\frac{d}{dn_{eff}}[\frac{2}{\pi}N(a)]$, which we found, is not much different in the index ($n_{eff,m}$) range we adresse (i.e. $n_b$>3). Based on Equ. (S34), frequency spacings in different configurations are investigated and presented in Fig. S3. Compared to a step-index waveguide with $n_b$=3 and $a$=350nm, the frequency spacings of linear-shape graded-index waveguide presents a faster change but in a basically similar line-shape. Even though using a wider graded-index waveguide with $a$=500nm or smaller boundary index $n_b$=2.6, the lineshapes in the concerned range of effective index ($n_{eff} < n_b$, i.e. [1.6, 3] and [1.6, 2.6], respectively) still behave monotonously. The only thing that matters is the absolute value frequency spacing. This result unambiguously confirms us that, without the assistance of a variable "effective width" supported by the self-adaptive boundary, there is no way to reshuffle the frequency spacings in the multi-mode scheme, since the integral range has a significant effect on the left part of eigen dispersion equation.

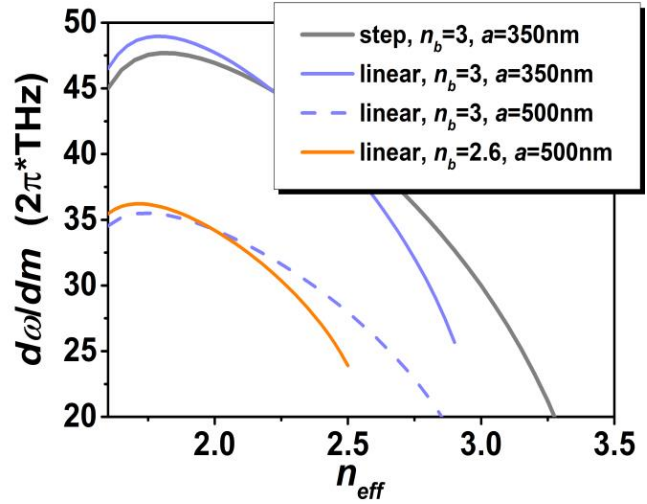

**Fig. S4.** Analytical Frequency spacings as a function of effective index $n_{eff}$, with different *order* and *a* values, using the classical condition $n_{eff} < n_b$.

**REFERENCES**

(1) K. Okamoto, "Fundamentals of Optical Waveguides," Elsevier 2006.